%% file: main.tex
\documentclass[10pt,twocolumn,letterpaper]{article}

 \usepackage{cvpr}              


\definecolor{cvprblue}{rgb}{0.21,0.49,0.74}
\usepackage[pagebackref,breaklinks,colorlinks,allcolors=cvprblue]{hyperref}
\usepackage{threeparttable}
\def\paperID{10317} 
\def\confName{CVPR}
\def\confYear{2025}

\title{Neural Video Compression with Context Modulation}

\author{Chuanbo Tang\quad Zhuoyuan Li\quad Yifan Bian\quad Li Li\quad Dong Liu\thanks{
\emph{Corresponding author: Dong Liu.}}\\
{\small MOE Key Laboratory of Brain-Inspired Intelligent Perception and Cognition}\\
{\small University of Science and Technology of China, Hefei 230027, China}\\
{\tt\small \{cbtang,zhuoyuanli\}@mail.ustc.edu.cn, togelbian@gmail.com, \{lil1,dongeliu\}@ustc.edu.cn}}

\begin{document}
	\maketitle
	\begin{abstract}
		Efficient video coding is highly dependent on exploiting the temporal redundancy, which is usually achieved by extracting and leveraging the temporal context in the emerging conditional coding-based neural video codec (NVC). Although the latest NVC has achieved remarkable progress in improving the compression performance, the inherent temporal context propagation mechanism lacks the ability to sufficiently leverage the reference information, limiting further improvement. In this paper, we address the limitation by modulating the temporal context with the reference frame in two steps. Specifically, we first propose the flow orientation to mine the inter-correlation between the reference frame and prediction frame for generating the additional oriented temporal context. Moreover, we introduce the context compensation to leverage the oriented context to modulate the propagated temporal context generated from the propagated reference feature. Through the synergy mechanism and decoupling loss supervision, the irrelevant propagated information can be effectively eliminated to ensure better context modeling. Experimental results demonstrate that our codec achieves on average 22.7\% bitrate reduction over the advanced traditional video codec H.266/VVC, and offers an average 10.1\% bitrate saving over the previous state-of-the-art NVC DCVC-FM. The code is available at \url{https://github.com/Austin4USTC/DCMVC}.
	\end{abstract}
	

	\section{Introduction}
	
	Temporal prediction~\cite{li2022global,li2024uniformly,li2024object,dong2023temporal} plays an essential role in removing temporal redundancy for video compression. Since DCVC~\cite{li2021deep} shifted the paradigm from traditional residual coding to conditional coding, temporal prediction has been represented efficiently by the temporal context for the emerging conditional coding-based neural video compression framework~\cite{li2021deep,sheng2022temporal,li2022hybrid,ho2022canf,li2023neural,qi2023motion,li2024neural,tang2024offline}. The temporal context is regarded as the condition to encode and decode the current frame, so generating high-quality temporal context is the key to achieve a better compression performance.
	
	\begin{figure}
		\begin{center}
			\includegraphics[width=\linewidth]{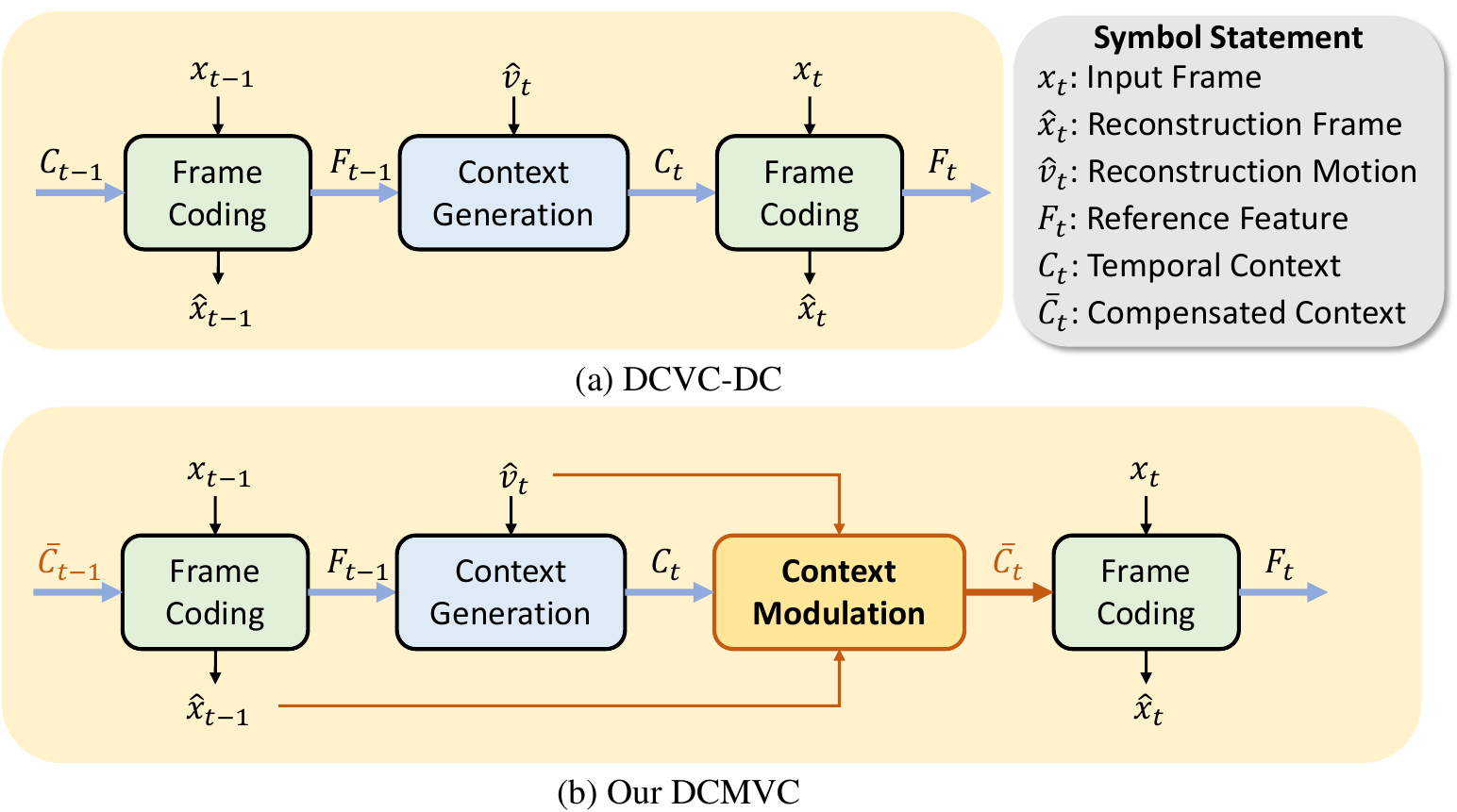}
		\end{center}
	    \vspace{-1em}
		\caption{Comparison of context generation for our DCMVC (Deep Context Modulation for Video Compression) with previous state-of-the-art compression scheme, DCVC-DC.}
		\label{fig:compare_frame}
		\vspace{-1em}
	\end{figure}
	
	Recent conditional coding-based neural video codecs (NVCs) have invested significant efforts in better leveraging the temporal context to realize effective inter-frame prediction. Based on DCVC~\cite{li2021deep}, DCVC-TCM~\cite{sheng2022temporal} further mines temporal context from the propagated reference feature and motion vector (MV), and the reference propagation structure mechanism is widely adopted in the following works~\cite{li2022hybrid,li2023neural,qi2023motion,li2024neural,sheng2024spatial}. 
	As shown in Fig.~\ref{fig:compare_frame} (a), taking our baseline DCVC-DC~\cite{li2023neural} as an example, the context of current frame ${C}_{t}$ is generated through the context generation mechanism with the input of propagated adjacent reference feature ${F}_{t-1}$ and reconstructed MV $\hat{v}_t$. Based on the DCVC-DC, considering temporal error accumulation problem in long prediction chain, DCVC-FM~\cite{li2024neural} has proposed a periodically refreshing mechanism in the process of context generation, which manually switches the input of context generation from the propagated reference feature ${F}_{t-1}$ to reference frame $\hat{x}_{t-1}$. Through the modulation on the temporal feature under a fixed temporal period, DCVC-FM mitigates the error propagation and maintains a good inter-frame prediction quality in long prediction chain.

	
	Despite the progress made by DCVC-FM, we rethink the temporal context modeling in the conditional coding-based framework. Motivated by the inherent reference propagation structure mechanism and benefits of periodically refreshing mechanism in DCVC-FM, we find the propagated reference feature conveys irrelevant information in the prediction chain, though it contains more information than reference frame. Simultaneously, switching between reference frames and reference features with the fixed temporal period may not effectively exploit the reference information.
	
	To address this issue, we propose the context modulation to generate a high-quality temporal context exploiting the reference information in both pixel and feature domain shown in Fig.~\ref{fig:compare_frame} (b). In this paper, we propose the two-step context modulation framework. First, we consider how to leverage the reference frame to learn an additional temporal context to assist the propagated temporal context generated from the propagated reference feature. Specifically, in order to adequately exploit the reference information in the pixel domain, we propose the flow orientation to extract the oriented flow between the reference frame and prediction frame. Then, we learn the oriented context from the reference frame and oriented flow for further modulating the propagated context. Second, given the propagated context and oriented context, we investigate how to manage them for temporal context modeling, and propose the context compensation to modulate the propagated context with the assistance of oriented context. To further facilitate the synergy mechanism, we also design a training constraint in the form of decoupling loss function promoting better complementarity of two contexts during the training. Through the context compensation, the irrelevant propagated information is removed to ensure better context modeling, which is also beneficial in alleviating error propagation.
	
	With the aforementioned flow orientation and context compensation methods, we develop a novel NVC termed DCMVC, which stands for Deep Context Modulation for Video Compression. Experiments demonstrate that our DCMVC surpasses previous SOTA neural video compression schemes under both intra-period settings of 32 and -1. Compared to the advanced traditional codec H.266/VVC, our model achieves on average 22.7\% bitrate saving on the tested videos. Our model also outperforms the best neural video codec DCVC-FM by 10.1\% bitrate reduction, showing the effectiveness of our proposed scheme. 
	
	Our contribution can be summarized as follows:
	
	
	\begin{itemize}
		\item We propose the context modulation in the conditional coding-based framework to generate high-quality temporal context exploiting the reference information in both pixel and feature domain. Our model can enable better temporal context and alleviate error propagation.
		
		\item We propose the flow orientation method to mine inter-frame correlation between the reference frame and prediction frame. It enables our DCMVC to generate additional oriented temporal context from the reference frame.   
		
		\item We propose the context compensation method to modulate the propagated temporal context with oriented context by using synergy mechanism and decoupling loss. It eliminates the irrelevant propagated information to ensure better context modeling.
		
		
		\item Our DCMVC can achieve 22.7\% bitrate reduction over the traditional video codec H.266/VVC, and obtain on average 10.1\% bitrate saving compared to the previous SOTA neural video codec DCVC-FM.
		
	\end{itemize}

	\section{Related Work}
	\subsection{Neural Video Compression}
	With the fast development of neural image compression~\cite{balle2016end, balle2018variational,cheng2020image, minnen2020channel, he2021checkerboard, he2022elic, liu2023learned,zhang2024practical,zhang2024generalized,liu2025region}, neural video compression has drawn increasing attention in recent years. Existing NVCs can be roughly classified into four classes: volume coding-based video codecs~\cite{Habibian_2019_ICCV,sun2020high,pessoa2020end}, implicit neural representation-based codecs~\cite{chen2021nerv,chen2023hnerv,zhao2023dnerv,kim2023c3,kwan2024hinerv}, residual coding-based video codecs~\cite{lu2019dvc,lu2020content,lu2020end, lin2020m, hu2020improving,yang2020learning, agustsson2020scale,hu2021fvc,shi2022alphavc,hu2022coarse,hu2023complexity,li2023high, chen2024group}, and conditional coding-based video codecs~\cite{li2021deep,sheng2022temporal,li2022hybrid,ho2022canf,li2023neural,qi2023motion,li2024neural,sheng2024spatial,tang2024offline,jiang2024ecvc,sheng2024nvc}. The first two categories of compression schemes still have considerable room for improvement in terms of compression performance, while the latter two are currently the dominant solutions and have already achieved impressive progress. Residual coding-based video codecs leverage the optical flow estimation network~\cite{ranjan2017optical,sun2018pwc,teed2020raft} to generate the prediction frame, then its residual with the current frame is encoded to remove the temporal redundancy. Different from the residual-coding paradigm, the emerging conditional coding-based video codecs learn the temporal context in more flexible ways and have an entropy bound lower than or equal to residual coding. Our scheme adopts the conditional coding-based framework and makes improvements on temporal context learning.

	
	\subsection{Temporal Context Learning}
	In NVC, the temporal context is usually extracted from the previous reconstruction frames and the current frame, then utilized to assist the current frame coding. For residual coding-based scheme, DVC~\cite{lu2019dvc} utilized the optical flows and reference frame to generate the prediction frame as the temporal context. M-LVC~\cite{lin2020m} further improved the context modeling with multiple reference frames. To improve the feature alignment, FVC~\cite{hu2021fvc} adopted the deformable convolution~\cite{dai2017deformable} to obtain more accurate context in the feature domain. AlphaVC~\cite{shi2022alphavc} proposed pixel-to-feature motion prediction to improve the accuracy of temporal context modeling at the encoder side.
	
	For conditional coding-based schemes, DCVC~\cite{li2021deep} has proposed to generate the context in feature domain to remove the temporal redundancy rather than relying on the subtraction-based residual in pixel domain. The subsequent DCVC series~\cite{sheng2022temporal,li2022hybrid,li2023neural,qi2023motion,li2024neural} have focused on enhancing the temporal context learning for achieving better compression performance. DCVC-TCM~\cite{sheng2022temporal} proposed the reference feature propagation mechanism to learn the multi-scale temporal context, which is widely adopted in the following works. DCVC-HEM~\cite{li2022hybrid} further proposed the hybrid entropy model to better capture the temporal and spatial dependency lies in the context. Group-based offset diversity~\cite{li2023neural} and hybrid context generation module~\cite{qi2023motion} were proposed to improve the quality of the temporal context. The latest DCVC-FM~\cite{li2024neural} alleviated the error propagation through modulating the temporal feature. The reference frame was periodically selected to generate the context rather than from propagated reference features for stopping the quality degradation in long prediction chain. 
	
	Although these works have achieved progressive compression performance, mining high-quality temporal context remains exploration. The inherent reference propagation structure mechanism leads to quality degradation in long prediction chain, because the propagated reference feature may contain more irrelevant information than reference frame. 
	Although DCVC-FM~\cite{li2024neural} mitigates error propagation by periodically switching the reference frame in long prediction chains, only this manual context modeling mechanism may not sufficiently exploit the reference information. In this paper, we address the limitation by proposing context modulation to sufficiently exploit the reference information in both pixel and feature domain.
	

	\begin{figure}
		\begin{center}
			\includegraphics[width=\linewidth]{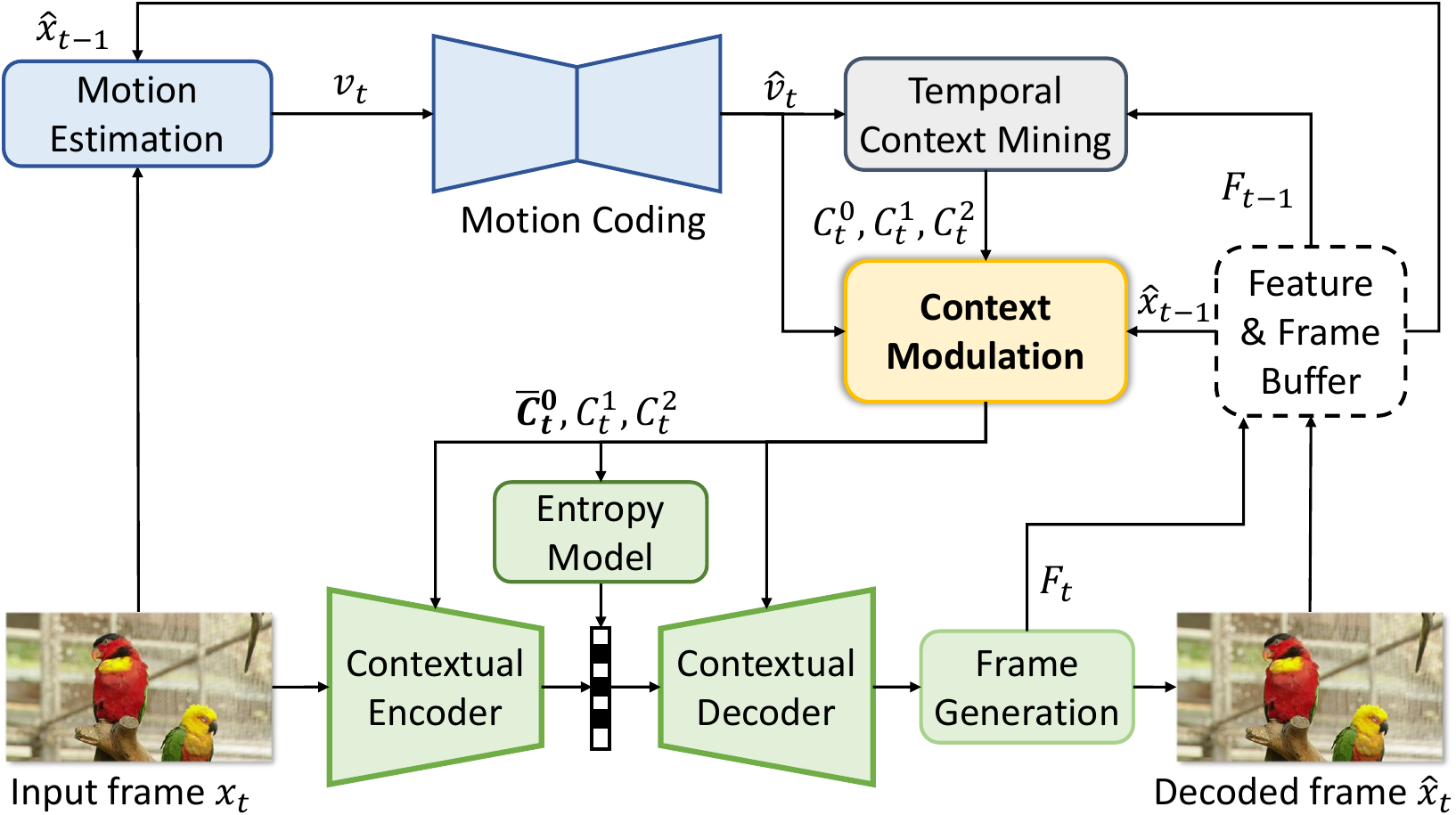}
		\end{center}
		\vspace{-1.1em}
		\caption{Our DCMVC framework. Based on the contextual-coding framework, we propose the context modulation to generate the compensated temporal context $\overline{C}_t^0$ with the input of propagated temporal context $C_t^0$, decoded flow $\hat{v}_{t}$, and reference frame $\hat{x}_{t-1}$.}
		\label{fig:overview.pdf}
		\vspace{-1.1em}
	\end{figure}
	\section{Proposed Method}
	\subsection{Overview}
	Our DCMVC is built on the emerging conditional coding-based framework, DCVC-DC~\cite{li2023neural}. In Fig.~\ref{fig:overview.pdf}, our proposed framework DCMVC is illustrated. First, the current frame $x_t$ and reference frame $\hat{x}_{t-1}$ (the previous decoded frame) are input into the motion estimation for generating the MV $v_t$, which is represented by optical flows. The estimated flow $v_t$ is compressed through the motion coding module. Second, with the decoded flow $\hat{v}_{t}$ and the propagated reference feature $F_{t-1}$, the temporal context mining is performed to learn multi-scale propagated temporal contexts $C_t^0$, $C_t^1$, $C_t^2$. To better model temporal context, we propose the context modulation to generate the largest-scale compensated temporal context $\overline{C}_t^0$ given the reference frame $\hat{x}_{t-1}$ and the decoded flow $\hat{v}_{t}$. 
	The compensated temporal context $\overline{C}_t^0$, along with $C_t^1$ and $C_t^2$, serves as conditions for the contextual encoder, entropy model, and decoder. Third, given the feature decoded by the contextual decoder, the reconstruction frame $\hat{x}_{t}$ and reconstruction feature ${F}_{t}$ are generated by the frame generator, and ${F}_{t}$ is propagated as the reference of next frame coding. 
	
\begin{figure*}
		\begin{center}
			\includegraphics[width=0.9\linewidth]{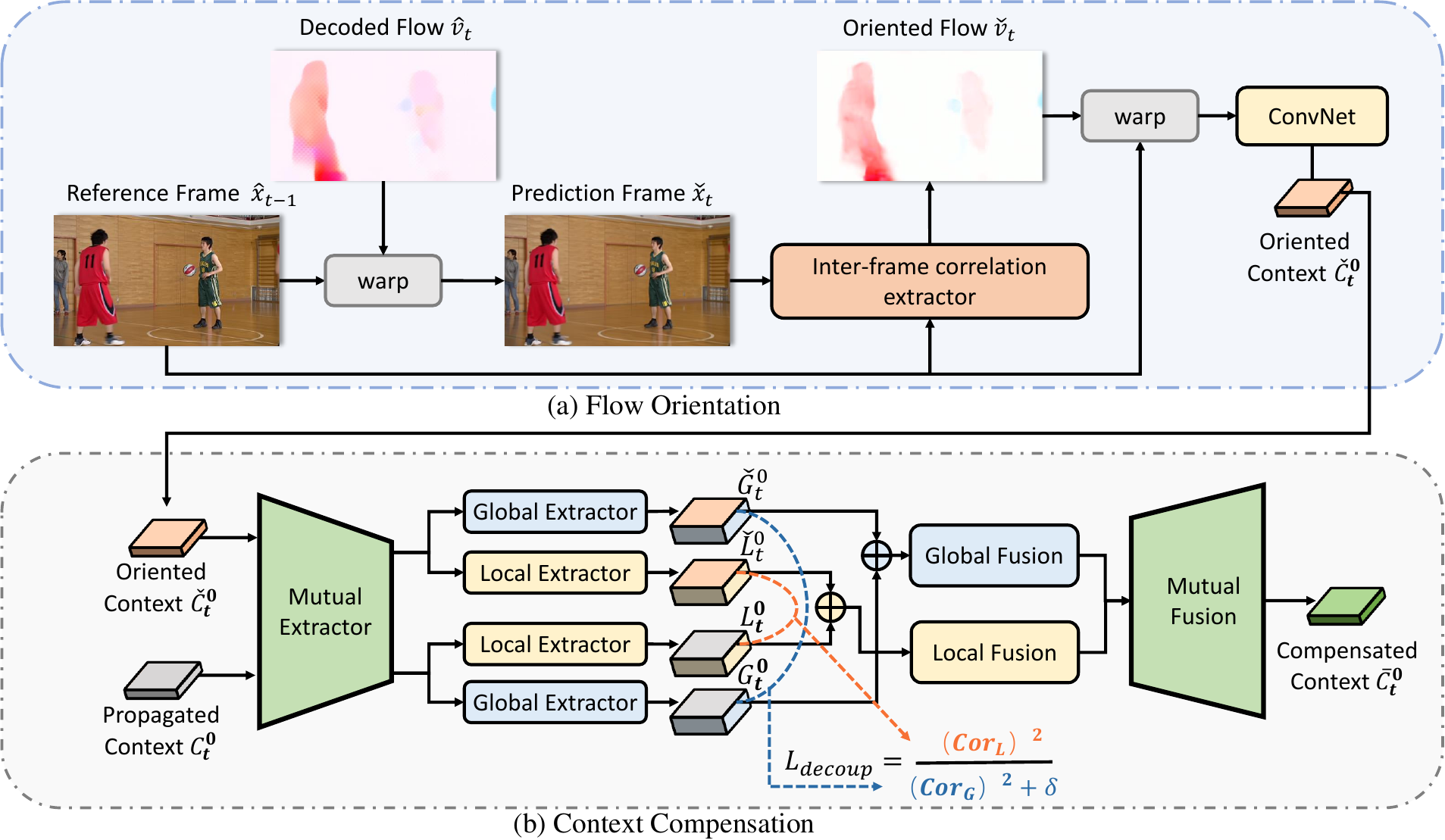}
		\end{center}
		\caption{(a) Framework overview of the flow orientation in context modulation. $\check{C}_t^0$ is the oriented context generated from reference frame $\hat{x}_{t-1}$ and oriented flow $\check{v}_{t}$. (b) Framework overview of the context compensation in context modulation. ${C}_t^0$ is the context generated through propagated reference feature $F_{t-1}$ and decoded flow $\hat{v}_{t}$. $\check{G}_t^0$ and ${G}_t^0$ are global features extracted from $\check{C}_t^0$ and ${C}_t^0$, respectively, and ${Cor}_G$ is the cosine correlation between $\check{G}_t^0$ and ${G}_t^0$. Similarly, $\check{L}_t^0$ and ${L}_t^0$ are local features extracted from $\check{C}_t^0$ and ${C}_t^0$, respectively, and ${Cor}_L$ is the cosine correlation between $\check{L}_t^0$ and ${L}_t^0$. The decoupling loss function ${L}_{decouple}$ consists of the cosine similarity of both global features and local features. $\overline{C}_t^0$ is the context generated from our proposed context compensation.}
		\label{fig:structure_detail}
	\end{figure*}

	Our proposed context modulation consists of flow orientation and context compensation, with workflows shown in Fig.~\ref{fig:structure_detail}, and detailed in Section~\ref{sec:flow_ori} and Section~\ref{sec:CC}.
	
	\subsection{Flow Orientation}\label{sec:flow_ori}
	
	Previous conditional coding-based NVCs encode and decode the current frame with the condition of the propagated temporal context. The propagated context is inherently generated through the decoded flow $\hat{v}_t$ and propagated reference feature ${F}_{t-1}$. Inspired by the error propagation phenomenon of the inherent reference propagation structure mechanism in DCVC-DC, observed in~\cite{li2024neural}, the propagated reference feature brings irrelevant information impacting error accumulation in the prediction chain. In contrast, the reference frame $\hat{x}_{t-1}$, which is the adjacent reconstructed frame, is constrained by the distortion loss in the rate-distortion (RD) loss function, so the reference frame contains as little irrelevant information as possible compared to the propagated reference feature. To sufficiently exploit the reference frame to generate an additional temporal context, we propose the flow orientation method.
	
	Fig.~\ref{fig:structure_detail} (a) shows the generation of the oriented temporal context $\check{C}_t^0$ in the flow orientation. With the decoded flow $\hat{v}_{t}$ parsed from the motion decoder, the prediction frame $\check{x}_t$ is obtained by warping the reference frame $\hat{x}_{t}$. Due to the RD constraint of motion bits and prediction quality in training, the representation capacity of the prediction frame for the generation of temporal context is limited. Therefore, we propose to introduce an additional extraction module to extract the inter-frame correlation for remedy. Specifically, the Spynet~\cite{ranjan2017optical} is used as a pyramid inter-frame correlation extractor to search the temporal correlation information between the reference frame and prediction frame, which is named as the oriented flow $\check{v}_t$. 
	As shown in Fig.~\ref{fig:flow_compare} (a), we visualize the estimated flow $v_t$ (estimated from the reference frame and current frame), decoded flow $\hat{v}_{t}$ (decoded from the motion decoder), oriented flow $\check{v}_t$, and their corresponding warp frames. 
	Due to the RD constraint of motion bits, the uncertainty of the estimated flow and decoded flow may incur prediction errors in the warp frames. In contrast, the oriented warp frame generated by oriented flow $\check{v}_t$ achieves better prediction results than those generated by $v_t$ and $\hat{v}_{t}$, which indicates that the oriented flow has learned additional temporal correlation information for the temporal context mining. With the guidance of oriented flow $\check{v}_t$, the reference frame $\hat{x}_{t}$ is aligned to extract the oriented temporal context $\check{C}_t^0$. Unlike traditional codecs, leveraging the training manner of NVC, the oriented temporal context enables the network to recover critical temporal correlation loss without extra signal overhead.

	
	\subsection{Context Compensation}\label{sec:CC}
	After obtaining the oriented temporal context and the propagated temporal context from different reference mechanisms, we consider how to effectively manage these temporal contexts for better temporal context modeling. To achieve the synergy of two kinds of temporal contexts, we propose the context compensation method to modulate the propagated context with oriented context.   
	
	Motivated by the good traditions of multi-source perception in~\cite{ma2019deep, zhao2023cddfuse,li2024spiking,zhao2024equivariant}, we design a global-local context compensation network to enable better complementarity of the oriented temporal context and propagated temporal context. Besides, decoupling loss is designed to supervise the network for facilitating better synergy of different contexts.
	
	Fig.~\ref{fig:structure_detail} (b) shows the workflow of our proposed context compensation. First, the oriented temporal context $\check{C}_t^0$ is generated from the flow orientation, and the largest-scale temporal context ${C}_t^0$ is generated from the temporal context mining, which is performed by the multi-scale alignment~\cite{li2023neural} of the propagated feature $F_{t}$ with decoded flow $\hat{v}_{t}$. Then, with the global-local synergy mechanism and the supervision of decoupling loss, the oriented context $\check{C}_t^0$ is used to modulate the propagated context ${C}_t^0$ for learning the compensated temporal context $\overline{C}_t^0$. In detail, the structure of context compensation can be divided into two parts: feature extraction and feature fusion. 
	
	{\bfseries Feature Extraction.} Specifically, the shallow features from $\check{C}_t^0$ and $C_t^0$ are learned by the shared mutual extractor. 
	Then, the private global extractors learn the global features from $\check{C}_t^0$ and $C_t^0$ , while the private local extractors learn the local features from $\check{C}_t^0$ and $C_t^0$. The two contexts $\check{C}_t^0$ and $C_t^0$ are derived from similar reference resources (reference frame and propagated reference features), so the global features of two contexts are more similar than local features. Therefore, we design the local extractor to extract and preserve as much detail information of two contexts as possible, and we adopt invertible neural networks with affine decoupling layers in both local extractor and local fusion.
	
	{\bfseries Feature Fusion.} After obtaining the global and local features from the oriented context $\check{C}_t^0$ ($\check{G}_t^0$ and $\check{L}_t^0$) and propagated context ${C}_t^0$ (${G}_t^0$ and ${L}_t^0$), the global features and local features of the two contexts are added and then input to the global and local fusion network, respectively. To maintain the performance by leveraging the consistency assumption, the networks of extractor and fusion adopt the same architecture. Then, the global and local fusion net are used to generate the fused global feature and local feature, which are further concatenated to input to the shared mutual fusion for learning the final compensated context $\overline{C}_t^0$. 
	
	\begin{figure}
		\begin{center}
			\includegraphics[width=\linewidth]{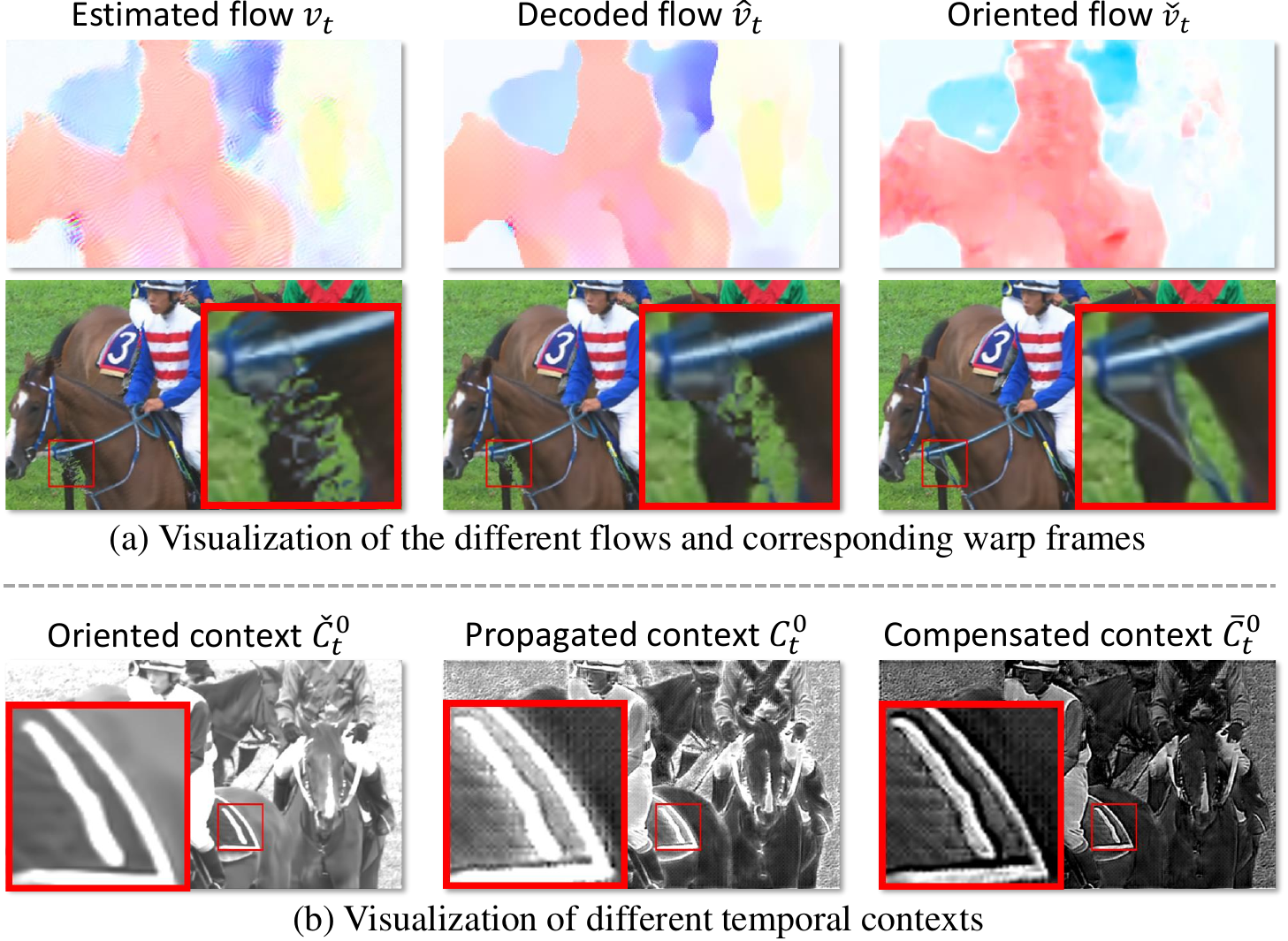}
		\end{center}
		\vspace{-0.2em}
		\caption{(a) Visualization of the estimated flow $v_t$ (estimated from reference frame and current frame), decoded flow $\hat{v}_{t}$ (decoded from motion decoder), oriented flow $\check{v}_t$ (generated from reference frame and prediction frame), and their warp frames. (b) Visualization of oriented context $\check{C}_t^0$, propagated context ${C}_t^0$, and compensated context $\overline{C}_t^0$ obtained from the context compensation.}
		\label{fig:flow_compare}
	\end{figure}
	
	
	As illustrated in Fig.~\ref{fig:flow_compare} (b), we visualize the oriented context $\check{C}_t^0$, propagated context ${C}_t^0$, and compensated context $\overline{C}_t^0$. Upon closer inspection of the contexts, we notice that the oriented context $\check{C}_t^0$ has smoother object edges, while the propagated context ${C}_t^0$ exhibits obvious prediction errors. With the synergy mechanism and guidance of decoupling loss, we can observe that the prediction errors are restored in compensated context $\overline{C}_t^0$. 
	
	{\bfseries Decoupling losses.} To facilitate the synergy mechanism~\cite{bai2024task,bai2024refusion} during the training, we design a decoupling loss $L_{decouple}$ shown as follows: 
	\begin{equation} 
		\label{com_loss}
		L_{decouple} = \frac{(Cor(\check{L}_t^0,{L}_t^0))^2}{(Cor(\check{G}_t^0,{G}_t^0))^2+\delta},
	\end{equation}
	where the $Cor(\cdot,\cdot)$ represents the cosine similarity operator. $\check{L}_t^0$ and ${L}_t^0$ denotes the local features of the oriented context and propagated context, and $\check{G}_t^0$ and ${G}_t^0$ denotes the global features of the oriented context and propagated context. $\delta$ indicates the smoothing term to ensure the values are positive, and is set to 1e-6. 
	
	Under the supervision of decoupling loss, the global information (e.g., structure and background) in both temporal contexts becomes more correlated, while the local information (e.g., texture and edges), being more specific to different contexts, tends to be less correlated. The ablation study in Section~\ref{sec:abl} shows the effectiveness of the decoupling loss and visualization of global and local features. 
	
	\subsection{Training Loss}\label{sec:training_loss}
	The video compression framework is trained in an end-to-end manner optimizing the loss function shown as follows:
	\begin{equation} 
		\label{rdloss}
		L = \lambda \cdot D + \alpha \cdot L_{decouple} + R,
	\end{equation}
	where $D$ represents the distortion between the input frame and the reconstructed frame. We use the Mean Squared Error (MSE) metric to measure the distortion. $\lambda$ and $\alpha$ are hyperparameters to control the weights of distortion and decoupling loss, respectively. $R$ denotes the number of bits used to encode the frame.
	
	
	\section{Experimental Results}
	\subsection{Experimental Settings}
	{\bfseries Datasets.} For training, we use the commonly used 7-frame Vimeo-90k~\cite{xue2019video} dataset to align with the training setting of the most existing NVCs, and all the videos are randomly cropped into $256 \times 256$ patches. Besides, we utilize the raw Vimeo videos\footnote{\url{http://toflow.csail.mit.edu}} to select a subset of 9000 sequences for 32-frame cascaded training, and all the videos are randomly cropped into $256 \times 384$ patches. For testing, we use UVG~\cite{mercat2020uvg}, MCL-JCV~\cite{wang2016mcl}, USTC-TD~\cite{li2024ustc}, and HEVC Class B, C, D, and E dataset~\cite{bossen2013common}. 
	\begin{table*}
		\caption{ BD-Rate (\%) comparison in RGB colorspace measured with PSNR. 96 frames with intra-period=32.}
		\vspace{-2.2em}
		\begin{center}
			\begin{threeparttable}
				\begin{tabular}{lcccccccc}
					\toprule
					& UVG  & MCL-JCV  &HEVC B &HEVC C &HEVC D &HEVC E &USTC-TD  & Average\\ \hline
					VTM    &0.0   &0.0       &0.0    &0.0    &0.0    &0.0  &{\bfseries0.0}   &0.0     \\ \hline
					DCVC~\cite{li2021deep} &133.9  &106.6 &119.6 &152.5  &110.9  &274.8 &139.6  &148.3   \\ \hline
					DCVC-TCM~\cite{sheng2022temporal} &23.1 &38.2 &32.8  &62.1  &29.0  &75.8 &75.3 &48.0   \\ \hline
					DCVC-HEM~\cite{li2022hybrid} &--17.2  &--1.6  &--0.7  &16.1 &--7.1 &20.7 &20.6 &4.4    \\ \hline
					SDD~\cite{sheng2024spatial}  &--19.7  &--7.1  &--13.7   &--2.3  &--24.9 &--8.4 &7.7 &--9.8  \\ \hline
					DCVC-DC~\cite{li2023neural} &--25.9  &--14.4 &--13.9 &--8.8 &--27.7 &--19.1 &10.8 &--14.1 \\ \hline
					DCVC-FM~\cite{li2024neural} &--20.4  &--8.1 &--10.3 &--8.4 &--25.8 &--21.9 &25.4 &--9.9    \\ \hline
					Our DCMVC&{\bfseries--30.6} &{\bfseries--17.3} &{\bfseries--14.5}    &{\bfseries--14.4}    &{\bfseries--31.6}  &{\bfseries--28.1} &0.4 &{\bfseries--19.4} \\
					\bottomrule
				\end{tabular}
				\begin{tablenotes}
					\item \footnotesize Note: The quality indexes of DCVC-FM are set as 36, 45, 54, 63 to match the bit-rate range of DCVC-DC.
				\end{tablenotes}
			\end{threeparttable}
		\end{center}
		\vspace{-1.5em}
		\label{table:ip32_psnr_TD}
	\end{table*}
	\begin{figure*}
		\begin{center}
			\includegraphics[width=\linewidth]{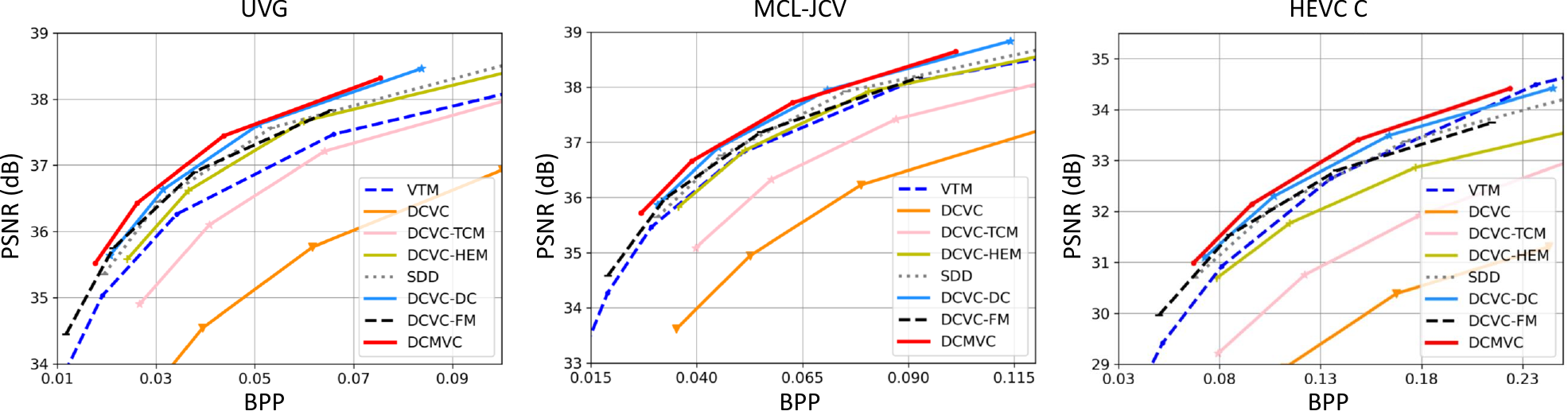}
		\end{center}
		\caption{Rate and distortion curve for UVG, MCL-JCV, and HEVC Class C datasets. The comparison is in RGB colorspace measured with PSNR, and the intra-period is set as 32. }
		\label{fig:rd_ip32.pdf}
		\vspace{-1.5em}
	\end{figure*}
	
	\begin{table*}
		\caption{ BD-Rate (\%) comparison in RGB colorspace measured with PSNR. 96 frames with intra-period=-1.}
		\vspace{-2em}
		\begin{center}
			\begin{threeparttable}
				\begin{tabular}{lcccccccc}
					\toprule
					& UVG  & MCL-JCV  &HEVC B &HEVC C &HEVC D &HEVC E &USTC-TD  & Average\\ \hline
					VTM    &0.0   &0.0       &0.0    &0.0    &0.0    &0.0     &{\bfseries0.0}  &0.0     \\ \hline
					DCVC~\cite{li2021deep} &259.5  &160.0 &212.2 &254.8  &180.6   &858.4 &168.2 &276.2   \\ \hline
					DCVC-TCM~\cite{sheng2022temporal} &61.6 &55.5 &61.7  &99.6  &50.4  &213.9 &90.2 &90.4   \\ \hline
					DCVC-HEM~\cite{li2022hybrid} &1.2  &4.9  &10.0  &30.0 &-1.1 &68.6 &27.2 &20.1    \\ \hline
					SDD~\cite{sheng2024spatial}  &--5.5  &--0.3  &--2.2   &16.9  &--18.2 &46.4 &12.3 &7.1  \\ \hline
					DCVC-DC~\cite{li2023neural} &--21.2  &--13.0 &--10.8 &--0.1 &--24.2 &--7.7 &11.9 &--9.3 \\ \hline
					DCVC-FM~\cite{li2024neural} &--24.3  &--12.5 &--11.7 &--8.2 &--28.5 &--26.6 &23.9 &--12.6    \\ \hline
					Our DCMVC&{\bfseries--35.6} &{\bfseries--22.8}    &{\bfseries--16.8}    &{\bfseries--14.8}  &{\bfseries--34.6} &{\bfseries--36.5} &1.9 &{\bfseries-22.7} \\
					\bottomrule
				\end{tabular}
				\begin{tablenotes}
					\item \footnotesize Note: The quality indexes of DCVC-FM are set as 36, 45, 54, 63 to match the bit-rate range of DCVC-DC.
				\end{tablenotes}
			\end{threeparttable}
		\end{center}
		\label{table:ip96_psnr_TD}
		\vspace{-1em}
	\end{table*}
	
	\begin{figure*}
		\begin{center}
			\includegraphics[width=\linewidth]{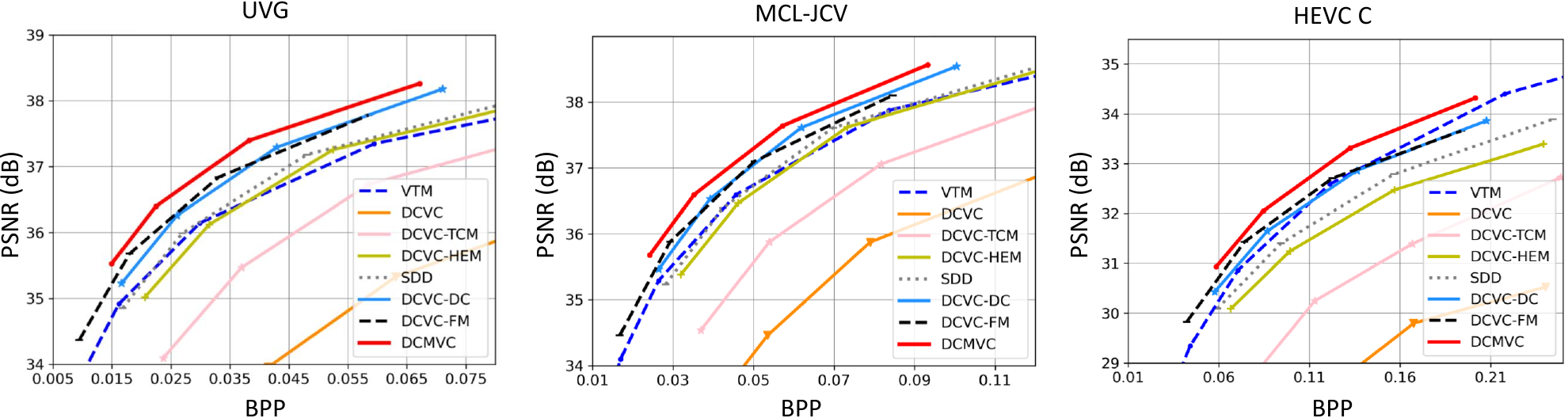}
		\end{center}
		\vspace{-1em}
		\caption{Rate and distortion curve for UVG, MCL-JCV, and HEVC Class C datasets. The comparison is in RGB colorspace measured with PSNR, and the intra-period is set as -1. }
		\label{fig:rd_ip96.pdf}
		\vspace{-0.5em}
	\end{figure*}
	
	
	{\bfseries Test Conditions.} Following most previous neural video compression schemes, we conduct our tests under low-delay settings. We test 96 frames for each video, and the intra-period is set to 32 and -1 for each dataset as two different settings. For evaluating the compression performance, we use BD-rate metric~\cite{bjontegaard2001calculation} to measure the compression ratio change. The negative numbers indicate bitrate saving, while positive numbers indicate bitrate increasing.
	
	We evaluate our scheme against two categories of codecs. For the traditional codec, we choose 
	VTM-13.2\footnote{\url{https://vcgit.hhi.fraunhofer.de/jvet/VVCSoftware_VTM/-/tree/VTM-13.2}} as our benchmarks, which is the official reference software of H.266/VVC~\cite{bross2021developments}. For neural video codec, we choose contextual coding-based NVCs as our benchmarks: DCVC~\cite{li2021deep}, DCVC-TCM~\cite{sheng2022temporal}, DCVC-HEM~\cite{li2022hybrid}, DCVC-DC~\cite{li2023neural}, SDD~\cite{sheng2024spatial}, and DCVC-FM~\cite{li2024neural}.
	
	{\bfseries Model Training.} We set four base $\lambda$ values (85, 170, 380, 840) to control the weight of distortion, and set $\alpha$ to 0.2 as the weight of the proposed decoupling loss $L_{decouple}$ in Eq.(\ref{rdloss}). Besides, in order to achieve the hierarchical quality structure, we follow~\cite{li2023neural} to set the hierarchical weight $w_t$ as (0.5, 1.2, 0.5, 0.9) to vary the base $\lambda$ in cycles during the training. Considering the CUDA memory pressure brought by the 32-frame cascaded training, Forward Recomputation Backpropagation (FRB)\footnote{\url{https://qywu.github.io/2019/05/22/explore-gradient-checkpointing.html}} is deployed during training on 4 NVIDIA A800 PCIe GPUs.
	
	\begin{figure}
		\begin{center}
			\includegraphics[width=\linewidth]{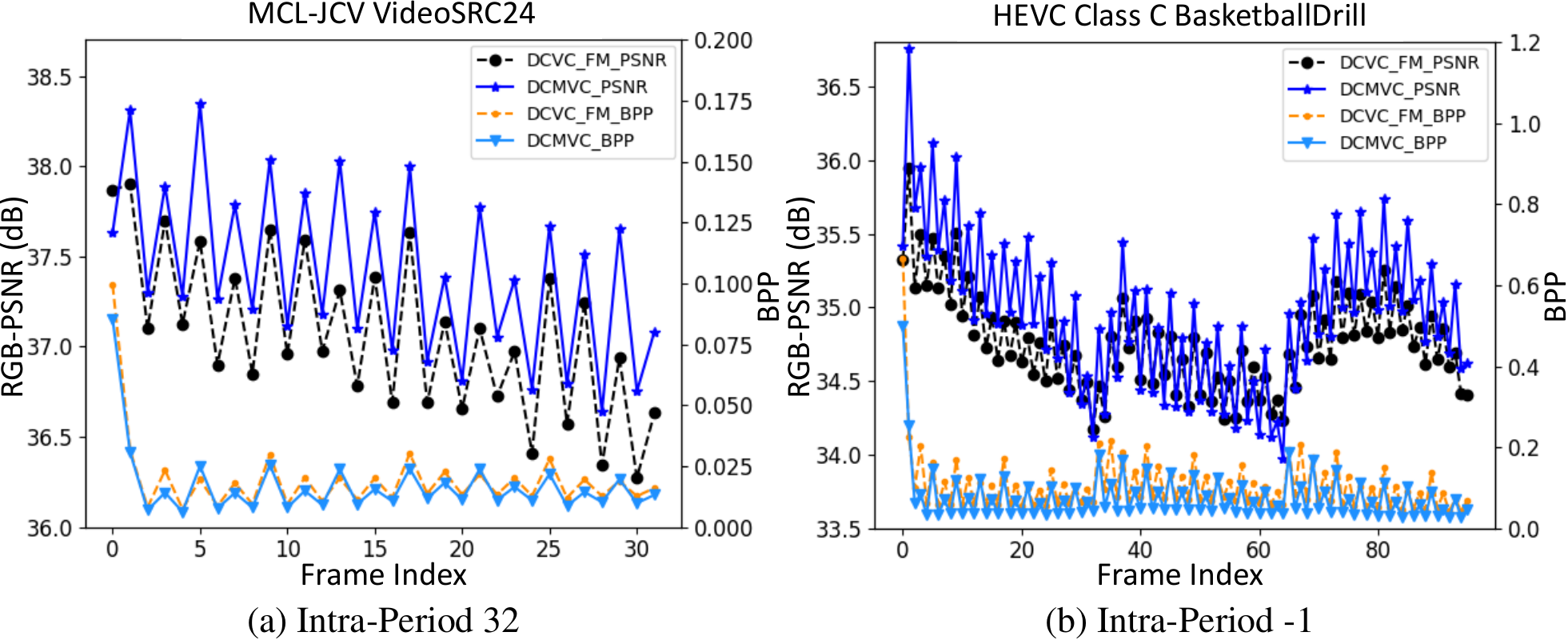}
		\end{center}
		\vspace{-1.5em}
		\caption{Quality and bitrate cost comparison across frames of DCVC-FM and our DCMVC.}
		\label{fig:err_pro}
		\vspace{-2em}
	\end{figure}
	
	\subsection{Comparisons with Previous SOTA Methods}
	Table~\ref{table:ip32_psnr_TD} reports the BD-Rate(\%) comparison results in terms of RGB-PSNR with intra-period set as 32 for 96 frames. When testing RGB videos, we convert the testing videos in YUV420 format to RGB format using FFmpeg. As shown in Table~\ref{table:ip32_psnr_TD}, compared to H.266/VVC, our DCMVC has achieved an average of 19.4\% bitrate saving. It also outperforms previous state-of-the-art (SOTA) neural video compression schemes DCVC-DC~\cite{li2023neural} (14.1\% bitrate saving), and DCVC-FM~\cite{li2024neural} (9.9\% bitrate saving) in each testing dataset. Fig.~\ref{fig:rd_ip32.pdf} shows the RD curves on testing datasets, where we can observe our DCMVC significantly surpasses previous SOTA schemes in RGB colorspace.
	
	We also focus on the evaluation under the intra-period setting of –1 to evaluate the model performance in longer prediction chain. Table~\ref{table:ip96_psnr_TD} illustrates the BD-Rate (\%) comparison results in terms of RGB-PSNR with intra-period set as –1. From the table, we can see our DCMVC also maintains the best RD performance compared with other benchmarks under both intra-period settings. Specifically, our scheme achieves an average of 22.7\% bitrate saving compared with H.266/VVC, and significantly performs better than the previous SOTA neural video compression scheme DCVC-FM~\cite{li2024neural} (12.6\% bitrate saving), which is also reported to be trained with 32-frame datasets. We provide the RD curves in Fig.~\ref{fig:rd_ip96.pdf}, where we can find our DCMVC brings obvious improvements compared to other benchmarks.
	
	As illustrated in Fig.~\ref{fig:err_pro}, we analyze the effectiveness of mitigating the error propagation in DCVC-FM and our DCMVC. In Fig.~\ref{fig:err_pro} (a), we provide the quality and bitrate cost across frames under the intra-period setting of 32. During the 32-frame temporal period, our DCMVC can better maintain the quality across frames with lower bitrate cost compared to DCVC-FM. Similarly, under the intra-period setting of -1, our DCMVC can better alleviate the quality degradation than DCVC-FM while costing fewer bits as shown in Fig.~\ref{fig:err_pro} (b). Observed from the aforementioned experimental results, it demonstrates that our context modulation performs effectively in both short intra-period and long intra-period measured with RD performances while alleviating the error propagation.
	
	\begin{table}
		\renewcommand\arraystretch{1.3}
		\caption{Ablation study on main techniques (\%)}
		\vspace{-2em}
		\begin{center}
			\scalebox{0.78}{
				\begin{tabular}{lccccccc}
					\toprule
					& $M_a$  & $M_b$     & $M_c$     &$M_d$      &$M_e$      &$M_f$      &$M_g$  \\ \hline
					Flow orientation        &        &\checkmark &           &\checkmark &\checkmark &           &\checkmark   \\ \hline
					Context compensation    &        &           &\checkmark &\checkmark &\checkmark &           &\checkmark    \\ \hline
					Decoupling loss         &        &           &           &           &\checkmark &           &\checkmark    \\ \hline
					Long-sequence training  &        &           &           &           &           &\checkmark &\checkmark     \\ \hline
					BD-Rate (\%)            &0.0     &-1.9       &-3.5       &-4.4       &-5.4       &-4.3       &-10.3 \\ 
					\bottomrule
				\end{tabular}
			}
		\end{center}
		\label{table:abl_bd}
		\vspace{-2em}
	\end{table}

	\subsection{Ablation Study}\label{sec:abl}
	To verify the effectiveness of each proposed method, we conduct ablation studies by evaluating the average RD performance on HEVC datasets with the intra-period set of 32. Table~\ref{table:abl_bd} illustrates the bitrate saving of each proposed method, where we can observe that each one achieves performance improvement by progressively adding each of them. We categorize our proposed methods into two main types based on the aspects of improvement: model architecture and optimization strategy.

	{\bfseries Model Structure.} In Table~\ref{table:abl_bd}, $M_a$ represents our reproduced baseline model DCVC-DC since its training code is not open-sourced. We first evaluate the two main techniques of model structure: flow orientation and context compensation. 
	When the model is equipped with the flow orientation alone ($M_b$), the oriented temporal context $\check{C}_t^0$ generated from the flow orientation and the propagated temporal context $C_t^0$ are concatenated directly, then the concatenated context is refined to generate the compensated context, which can reduce BD-rate by 1.9\% over $M_a$. To evaluate the effectiveness of the context compensation, we also design the $M_c$ model that the oriented temporal context $\check{C}_t^0$ is extracted from the prediction frame directly. Besides, we scale up the model scale of $M_c$ model to the final model ($M_d$) to exclude the performance improvement brought by the increase in model complexity. From Table~\ref{table:abl_bd}, we can find that $M_c$ model achieves 3.5\% bitrate saving compared to the baseline model $M_a$. In addition, when enabling flow orientation and context compensation together ($M_d$), there is a total 4.4\% BD-rate reduction over $M_a$. Observed the performance of $M_d$ model is better than $M_c$ model, we find that the extracted additional inter-frame correlation in flow orientation can provide a better oriented temporal context, which enables better temporal context modeling in the context compensation.

	
	\begin{figure}
		\begin{center}
			\includegraphics[width=\linewidth]{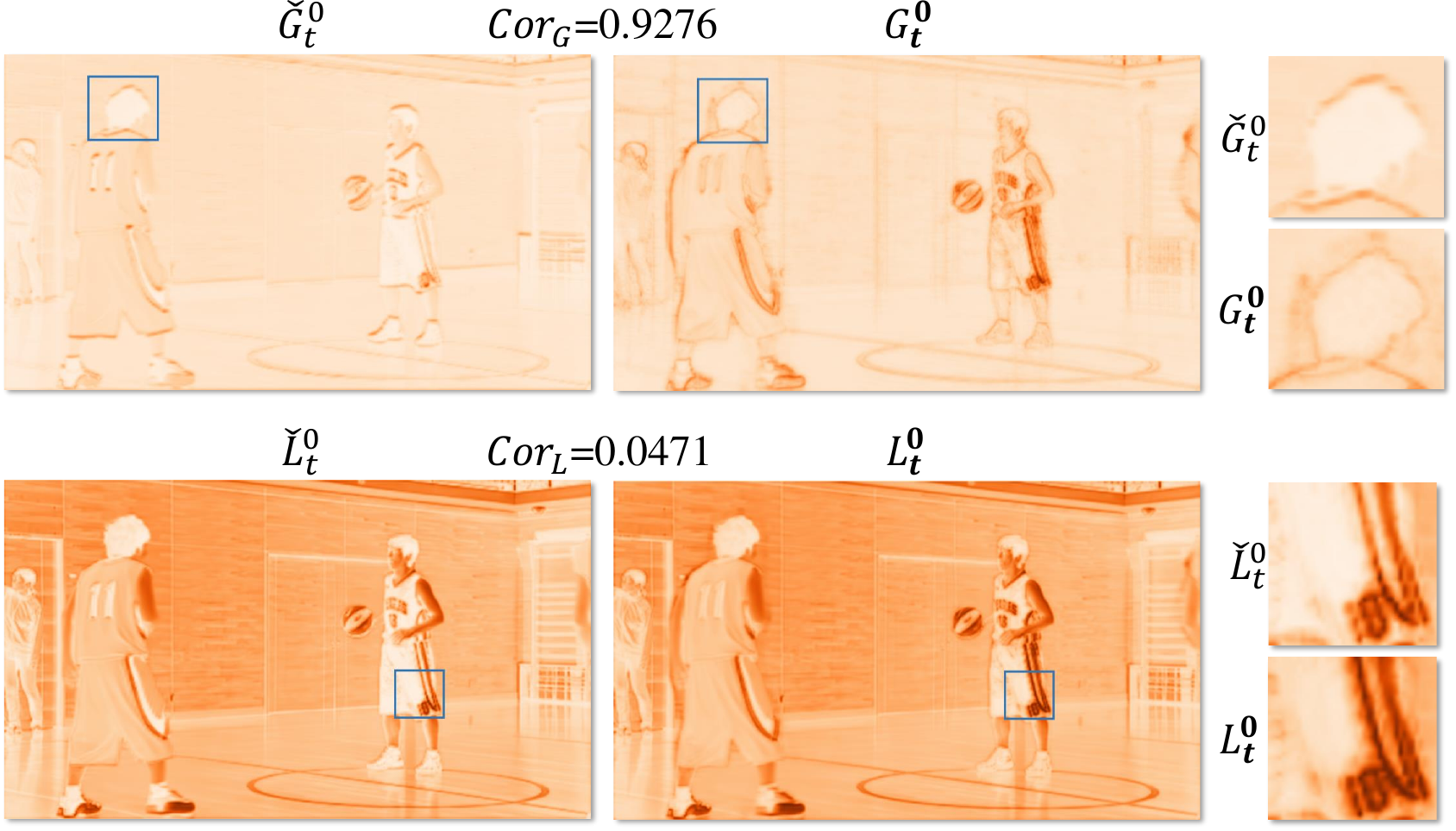}
		\end{center}
		\caption{Visualization of the global and local features extracted from the oriented context and propagated context.}
		\label{fig:context_cc}
		\vspace{-0.5em}
	\end{figure}
	{\bfseries Optimization Strategy.} Table~\ref{table:abl_bd} also shows the performance improvement achieved by implementing two enhancements in the training strategy without increasing model complexity. Adding the decoupling loss ($M_e$) into the objective function can not only enhance the interpretability of the model but also lead to a 1.0\% BD-rate reduction based on $M_d$ model. Fig.~\ref{fig:context_cc} illustrates the global and local features extracted from the oriented temporal context and propagated temporal context. First, we can obviously observe that the global extractors pay more attention to the background information, while the local extractors focus more on the texture information. Besides, with the supervision of decoupling loss, the global features are more correlated than the local features. Moreover, the visualization also shows that the global and local features extracted from the propagated temporal context contain more prediction errors than the features extracted from the oriented temporal context.
	Meanwhile, when applies longer sequences (32 frames) training solely ($M_f$) to our anchor $M_a$, it can reduce BD-rate by 4.3\%. Combined with the proposed context modulation, $M_g$ model can achieve a total 10.3\% bitrate saving over $M_a$. It confirms that providing richer reference information during long-sequence training enables the network to learn a more robust reference mechanism in temporal context modeling. 
	
	\begin{table}
		\renewcommand\arraystretch{1.4}
		\caption{Complexity comparison.}
		\vspace{-2em}
		\begin{center}
			\begin{threeparttable}
				\scalebox{0.75}{
					\begin{tabular}{lcccccc}
						\toprule
						& MACs    &Params & Encoding Time      &Decoding Time   \\ \hline
						DCVC-DC~\cite{li2023neural} &2764G    &19.78M&   663ms             &557ms                \\ \hline
						DCVC-FM~\cite{li2024neural} &2334G    &18.57M&   587ms             &495ms                \\ \hline
						SDD~\cite{sheng2024spatial} &3830G    &21.77M&   968ms             &775ms                \\ \hline
						Our DCMVC                   &4131G    &20.98M&   932ms             &810ms                \\ 
						\bottomrule
						
					\end{tabular}
				}
				\begin{tablenotes}
					\item \footnotesize Note: Tested on NVIDIA 3090 with using 1080p sequences as input.
				\end{tablenotes}
			\end{threeparttable}
		\end{center}
		\label{table:complexity}
		\vspace{-2em}
	\end{table}
	
	\subsection{Complexity Analysis}
	Table~\ref{table:complexity} presents the complexity comparison between our DCMVC and three latest SOTA NVCs: DCVC-DC~\cite{li2023neural}, DCVC-FM~\cite{li2024neural}, and SDD~\cite{sheng2024spatial}. We measure the model complexity with MACs, network parameters, encoding time, and decoding time. 
	The inference is conducted with 1080P frames on a NVIDIA 3090 GPU. Compared with DCVC-DC, SDD, and DCVC-FM, our DCMVC exhibits significantly higher RD performance at the cost of higher MACs and decoding time, while our encoding time is less than SDD. The moderate increase in MACs is mainly due to the complex computation in the largest-resolution temporal context. For the network parameters, our DCMVC has a limited increase in network parameters compared to DCVC-DC and DCVC-FM, while having fewer network parameters over SDD. 

	\section{Conclusion}
	In this paper, context modulation is proposed to address the limitation of inherent reference propagation structure mechanism in conditional coding-based NVCs. The proposed flow orientation leverages the reference frame to generate an additional oriented temporal context, then it is used for modulating the propagated temporal context in the proposed context compensation.
	Powered by our context modulation, DCMVC has further improved the compression ratio than previous SOTA NVCs, while mitigating the error propagation in the prediction chain. 
	
	Based on our exploration, investigating efficient temporal context modeling mechanisms still exhibits considerable potential. In temporal context generation, we still adopt rigid warp operation for alignment. For future work, learnable warp can be designed and utilized for more flexible temporal alignment. Besides, we only constrain MV indirectly by RD loss function during the training, which may be suboptimal for temporal dependency modeling. In the future, multiple predefined motion patterns can be learned as priors for efficient temporal context modeling. 
	
	
	{
		\small
		\bibliographystyle{ieeenat_fullname}
		\bibliography{main}
	}
	
	
\end{document}